\documentclass[apj]{emulateapj} 

\usepackage{times,amsmath,amssymb,amstext,graphicx,epstopdf,hyperref,subfigure,epsfig,aas_macros,natbib}

\usepackage{color}
\definecolor{myColor}{rgb}{0.9,0.9,0.9}  
\begin{document}
\renewcommand\bottomfraction{.9}
\title{Atmospheric Compositions of Three Brown Dwarfs and Implications for their Formation Conditions}
\shorttitle{Atmospheric Retrieval of Three Brown Dwarfs } 
\author{Nikku Madhusudhan\altaffilmark{1}, D\'aniel Apai \altaffilmark{2,3}, Siddharth Gandhi \altaffilmark{1}}
\altaffiltext{1}{Institute of Astronomy, University of Cambridge, Cambridge CB3 0HA, United Kingdom {\tt nmadhu@ast.cam.ac.uk}}
\altaffiltext{2}{Steward Observatory, and Lunar and Planetary Laboratory, The University of Arizona, 1640 E. University Blvd., Tucson, AZ 85718, USA}
\altaffiltext{3}{Earths in Other Solar Systems Team, NASA Nexus for Exoplanet System Science}

\begin{abstract}
The formation mechanisms and chemical compositions of brown dwarfs likely span a diverse range. If they are formed predominantly in isolation by gravitational collapse like dwarf stars then their compositions might follow those of dwarf stars, spanning a relatively narrow range in metallicities ($\sim$-0.4 to +0.4 dex) and predominantly oxygen-rich (C/O $\lesssim$ 0.7). On the other hand, giant planets in the solar system are all super-solar in metallicity ([C/H] $>$ 0.5~dex), which is thought to be a consequence of formation by core-accretion in a circumstellar disk. In this study we determine the atmospheric chemical compositions of three brown dwarfs and find them to be neither characteristic of dwarf stars nor giant planets. We derive high-precision atmospheric chemical abundances using high-S/N HST near-infrared spectra of three mid T Dwarfs which, together with two previously studied objects, display distinctly sub-solar metallicities and span C/O ratios of 0.4-1.0. These abundance patterns indicate either an old sub-stellar population and a multitude of formation environments or objects formed in circumstellar disks followed by ejection later.  We show that for a simple synthetic model population of brown dwarfs with the same spectral-type as our sample the predicted age distribution is predominantly older than the Sun and, hence, more metal poor, consistent with our derived abundances. The diverse C/O ratios we find are indicative of either different chemical reservoirs in their formation environments or different formation pathways. Our results open the possibility that cool brown dwarfs may provide important probes of early galactic chemical evolution and inhomogeneity. 
\end{abstract} 


\section{Introduction}
\label{sec:intro}

We are now entering a new era in characterizing exoplanets and sub-stellar objects. Going beyond object detections, the field is now moving towards detailed characterization of the atmospheres of exoplanets \citep{madhusudhan2014a,madhusudhan2016} and brown dwarfs \citep{apai2013,helling2014rev} through spectroscopic observations. Encoded within a spectrum of an exoplanet or brown dwarf is information about the chemical composition and manifold physical processes in its atmosphere. State-of-the-art observations are now beginning to provide both the high precision and long spectral baseline required to place detailed constraints on the various physicochemical properties of these atmospheres. Such observations in recent years are already providing good constraints on the chemical compositions of exoplanetary atmospheres \citep[see e.g., review by][]{madhusudhan2016}. The most observed exoplanets to date are hot giant transiting exoplanets whose large scale-heights and high temperatures ($\sim$800-3000 K) make them particularly conducive to atmospheric observations. Chemical detections have also been reported for a few directly-imaged giant exoplanets \citep[e.g.,][]{konopacky2013,barman2015}. The atmospheric chemical compositions are in turn beginning to provide the first insights into the possible formation conditions and migration mechanisms of exoplanets \citep[e.g.][]{oberg2011,madhusudhan2014c}. 

Given their wide orbital separations directly-imaged giant planets bear several similarities to isolated brown dwarfs. Both classes of objects have H$_2$-rich atmospheres with negligible levels of stellar irradiation, meaning that the internal heat dominates their spectra. Consequently, the atmospheric chemical compositions and physical processes in both classes of objects are expected to be quite similar for similar effective temperatures. The similarities in atmospheric properties make brown dwarfs exquisite analogs to investigate atmospheric processes and chemical compositions in exoplanets at large orbital separations. While there are only a handful of directly-imaged planets known with high-quality spectra, comparable or better quality spectra have been observed for numerous brown dwarfs using ground-based facilities \citep[e.g.][]{burgasser2010}. More recently, very high precision spectra (at SNR$\sim$1000) have also been obtained for several brown dwarfs using the HST Wide Field Camera 3 (WFC3) spectrograph in the near-infrared \citep{apai2013,buenzli2013} leading to high-confidence detections of H$_2$O and detailed characterization of variability in these objects. Multi-wavelength datasets are also providing the long spectral baseline and high precision required to derive joint constraints on the chemical compositions, temperature profiles, and cloud coverage in these objects. 

With the advent of atmospheric retrieval techniques for exoplanets \citep{madhusudhan2009,madhusudhan2011a} there is now growing interest in using similar techniques for brown dwarf spectra to retrieve their atmospheric properties. Recently, \citet{line2014,line2015} reported constraints on the atmospheric chemical abundances, temperature profiles, and bulk parameters of two brown dwarfs using their ground-based spectra in the near infrared. Given that the data quality for brown dwarfs is substantially better than that for exoplanets the former objects offer a valuable opportunity to constrain their atmospheric properties in unprecedented detail, and also to verify the models and retrieval techniques. The chemical abundances in brown dwarf atmospheres could in turn provide new insights into their formation mechanisms. Our goal in the present work is to conduct such a detailed retrieval analysis for three mid T Dwarfs, which together with two previously published objects, provide a sample of five T Dwarfs for comparative chemical characterization. 

In what follows, we discuss the target selection in section~\ref{sec:targets} and the observations in section~\ref{sec:data}. The modeling and retrieval approach is presented in section~\ref{sec:method} and the results in section~\ref{sec:results}. We summarize our results and discuss the implications of the derived chemical compositions in section~\ref{sec:discussion}. 

\begin{deluxetable*}{c c c c c c c}
\tablewidth{\textwidth}
\tabletypesize{\scriptsize}
\tablecaption{Fundamental properties of the T dwarfs studied in this work \tablenotemark{a}}
\tablehead{\colhead{\#} & \colhead{Object} & \colhead{2MASS ID} & \colhead{IR/Opt Sp. Type} & \colhead{2MASS J} & \colhead{2MASS J--K} & \colhead{Refs.}}
\startdata
    1 &    2MASS~J0243-2453 & 2MASS J02431371--2453298 & T6 / --   & 15.38$\pm$0.05 &  0.17$\pm$0.18 &B06, B14 \\
    2 &    2MASS~J0559-1404 & 2MASS J05591914--1404488 & T4.5 / T5 & 13.80$\pm0.02$& 0.23$\pm$0.06 &  B06, B03  \\
    3 &    2MASS~J2339+1352 &2MASS J23391025+1352284 &  T5 / --     & 16.24$\pm$0.010 & 0.09$\pm$0.33  & B06\\
    4 &    Gl~570D & 2MASS J14571496-2121477 &   & 15.32$\pm$0.048 & 0.08$\pm$0.16 & \\
    5 &    HD~3651B & --                                     &  T7.5 / --&  16.16$\pm$0.03  & $-$0.71$\pm$0.06    & B10
\enddata
\tablenotetext{a}{}
\tablenotetext{}{{\bf References:} B03 -- \citet{burgasser2003}; B06 --  \citet{burgasser2006}; B10 --  \citet{burgasser2010}; B14 -- \citet{Buenzli2014}}
\label{tab:targets}
\end{deluxetable*}

\begin{figure}[t]
\centering
\includegraphics[width = 0.5\textwidth]{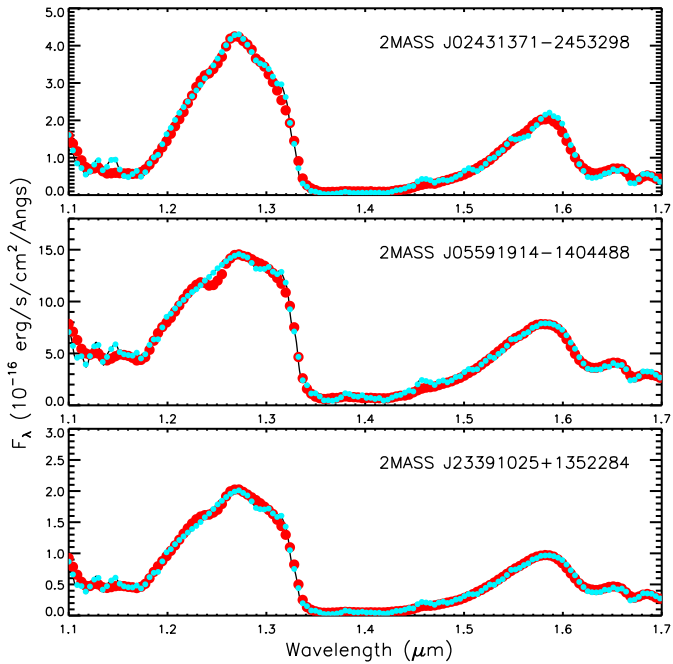} 
\caption{Observations and best-fit model spectra for the three targets in our study. The red circles with uncertainties show our HST WFC3 observations. The black curve shows the best-fit model spectrum, and the cyan circles show the model binned to the resolution of the data.} 
\label{fig:spectrum}
\end{figure}

\section{Target Selection}
\label{sec:targets}

In the present study we focus on a small number of prime targets. In order to test and validate our atmospheric abundance retrieval process and to explore the abundances of a few relatively simple brown dwarfs with very high quality and well-understood robust observations, we turned to the sample of brown dwarfs we have previously observed with the Hubble Space Telescope's Wide Field Camera 3 (HST/WFC3, \citealt[][]{MacKenty2010}) in its near-infrared grism model (G141 grism). Data in this mode has been collected for brown dwarfs in previous surveys \citep[][]{Buenzli2012,apai2013,Buenzli2014,Buenzli2015,Yang2015,Yang2016}, all of which have focused on searching for and characterizing brown dwarfs that display photometric variations due to the rotation of their heterogeneous atmospheres. Among the targets observed in these surveys for our study we searched for brown dwarfs based on the following criteria: a) presumably no or negligible visible cloud cover as predicted by atmospheric models, i.e. late T spectral type; b) well described by one-dimensional models, i.e. not showing temporal variability; c) probably single object (no evidence for companion in the high-resolution HST direct images); and, d) bright, i.e., has high signal-to-noise. Based on these criteria we identified up to seven brown dwarfs viable for this study; here we are presenting results for the first three (\object{2MASSJ02431371--2453298}, \object{2MASSJ05591914--1404488}, \object{2MASSJ23391025+1352284}). In addition to the three objects studied here, we also show results for two other T-type brown dwarfs (\object{HD 3651B} and \object{GJ 570D}) studied in \citet[][]{line2015} for comparison.

Table~\ref{tab:targets} provides an overview of the key properties of our targets.

\begin{figure*}[t]
\centering
\includegraphics[width = \textwidth]{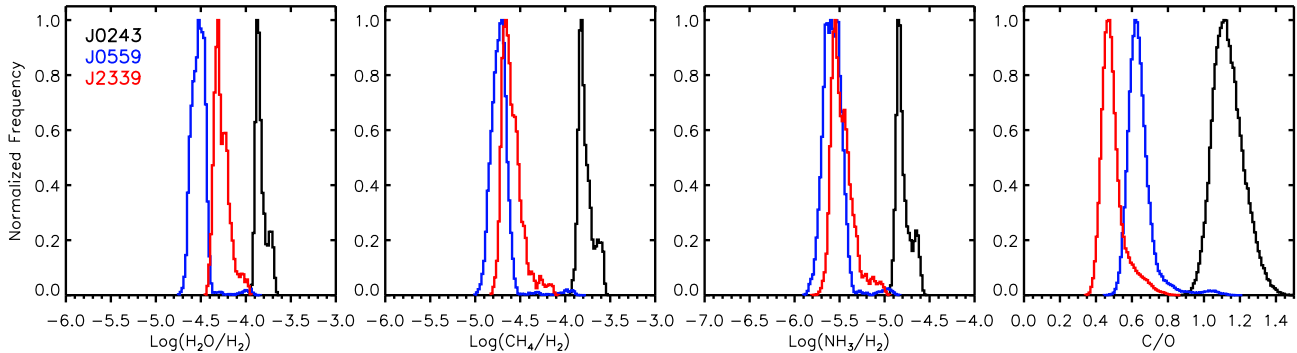} 
\caption{Posterior probability distributions of the molecular abundances of H$_2$O, CH$_4$, and NH$_3$ retrieved for  the three 2MASS T Dwarfs in our study.} 
\label{fig:histograms} 
\end{figure*}

\section{Observations and Comparison to Models}
\label{sec:data}
\subsection{Observations and Data Reduction}
The data used in this study have been obtained in our WFC3/G141 survey described in \citet[][]{Buenzli2014}. As the data acquisition, reduction, and calibration have been discussed in detail in that paper, we only highlight here the essential reduction steps, differences from the original reduction, and the uncertainty analysis as these are pertinent for the current study.

For each of our targets we used the 19 spectra taken with HST/WFC3/G141 during a single HST orbit. Our reduction is based on a combination of the aXe pipeline and a self-developed set of IDL/python routines; in summary, our steps include basic reduction (bad pixel and flat field corrections), wavelength calibration and re-mapping of the spectrum, and spectral extraction (for details see \citealt[][]{apai2013} and \citealt[][]{Buenzli2014}). In this study we also applied an additional step, aperture correction, to correct for the slight wavelength-dependent losses during the spectral extraction. While a negligible effect for the original studies which focused on temporal variations of the spectrum, the aperture correction allows a more accurate measurement of the apparent flux density.
We base our aperture correction on the relevant Instrument Status Report (WFC3 ISR 2011-05), which provides aperture corrections for a range of aperture sizes and wavelengths. We applied a linear interpolation in the aperture size dimension and a spline interpolation in the wavelength dimension to determine the aperture correction for our data and for each of our wavelength bins separately. As the final step, the spectra for each object were median-combined on an identically sampled  wavelength grid.

\subsection{Uncertainties}
The HST/WFC3/G141 observations provided highly repeatable spectra and with very high signal-to-noise. The stability of the instrument allows us to characterize the uncertainties beyond what is possible for most ground-based observations. 
Specifically, the uncertainties in our data set are dominated by two sources: random white noise -- primarily photon noise and read-out noise -- and systematic error affecting flux densities at all wavelengths -- primarily caused by slight changes in the instrument and telescope state. We characterize the uncertainties through the standard deviation of the 19 spectra in each wavelength bin. The standard deviation varies from a fraction of the percent in continuum bands (where the object is the brightest) to about 2\% of the flux density in water band (where the object is the faintest). We note that the standard deviation will not capture systematic changes that result, in effect, in slight changes of the efficiency of the instrument. While our analysis is largely insensitive to the precise absolute levels of the flux density, we estimate that such zero-point differences contribute at levels less than 0.5\% based on our analysis of identical datasets presented in \citep[][]{apai2013}. 

\subsection{Comparison to Models}
\label{sec:psf}
While the model calculations are performed at high spatial resolution, the HST/WFC3 G141 spectra is low resolution. Correctly comparing the models to the observations requires matching their resolutions. We do this by reducing the resolution of the models, i.e. by convolving the model spectrum with the HST point spread function. We modeled the point spread function using the Tiny Tim simulator \citep[][]{TinyTim2011}, which calculates a diffraction-limited point spread function considering the instrument, wavelength range, position on the detector, telescope focus, and other fundamental telescope parameters. We carried out our simulations for the F140W filter, which is a broad-band filter with its central wavelength approximately in the mid-range of our spectra. A comparison of the radial profile TinyTim provided to that measured perpendicular to the spectrum showed a very good match, verifying our approach. 

\begin{deluxetable*}{c c c c c c c}
\tablewidth{\textwidth}
\tabletypesize{\scriptsize}
\tablecaption{Atmospheric properties of T Dwarfs in our sample\tablenotemark{a}}
\tablehead{\colhead{\#} & \colhead{Object} & \colhead{Log(H$_2$O/H$_2$)} & \colhead{Log(CH$_4$/H$_2$)} & \colhead{Log(NH$_3$/H$_2$)} & \colhead{C/O\tablenotemark{b}} & \colhead{Log(g) (cgs)}}
\startdata
    1 &     2MASS~J0243-2453 & $-3.86_{-0.03}^{+0.10}$ & $-3.82_{-0.02}^{+0.14}$ & $-4.83_{-0.03}^{+0.14}$ & $ 0.87_{-0.02}^{+0.06}$ & $ 4.28_{-0.05}^{+0.25}$\\
    2 &     2MASS~J0559-1404 & $-4.49_{-0.12}^{+0.04}$ & $-4.71_{-0.11}^{+0.05}$ & $-5.58_{-0.10}^{+0.10}$ & $ 0.49_{-0.02}^{+0.20}$ & $ 3.44_{-0.30}^{+0.08}$\\
    3 &     2MASS~J2339+1352 & $-4.30_{-0.04}^{+0.12}$ & $-4.66_{-0.03}^{+0.17}$ & $-5.55_{-0.04}^{+0.19}$ & $ 0.35_{-0.21}^{+0.05}$ & $ 3.47_{-0.05}^{+0.31}$\\
    4 &     Gl~570D & $-3.24_{-0.08}^{+0.06}$ & $-3.14_{-0.09}^{+0.13}$ & $-4.50_{-0.11}^{+0.08}$ & $ 1.01_{-0.16}^{+0.13}$ & $ 5.12_{-0.17}^{+0.09}$\\ 
    5 &     HD~3651B & $-3.45_{-0.10}^{+0.10}$ & $-3.40_{-0.13}^{+0.13}$ & $-4.64_{-0.15}^{+0.15}$ & $ 0.90_{-0.14}^{+0.16}$ & $ 4.76_{-0.28}^{+0.27}$ 
\enddata
\tablenotetext{a}{The properties of the first three objects are retrieved in the present study. The properties of the remaining two objects are adopted from \citet{line2015}.}
\tablenotetext{b}{The C/O ratio is derived from the H$_2$O and CH$_4$ abundances, after allowing for 20\% of oxygen to be sequestered in silicates.}
\label{tab:results}
\end{deluxetable*}

\section{Atmospheric Modeling and Retrieval Method}
\label{sec:method}

In order to retrieve the atmospheric compositions and temperature profiles from spectra we use an atmospheric retrieval method that has been widely used for exoplanetary atmospheres \citep{madhusudhan2009,madhusudhan2011a}. The retrieval method comprises of a parametric forward model of an exoplanetary atmosphere coupled with a Bayesian parameter estimation algorithm to retrieve the model parameters (e.g., the chemical compositions, temperature profile, etc.) given a spectral dataset. The model computes line-by-line radiative transfer for a plane-parallel 1-D atmosphere with the assumptions of hydrostatic equilibrium as described in \citet{madhusudhan2009}. The composition and pressure-temperature ($P$-$T$) profile of the atmosphere are free parameters in the model. The model includes the major opacity sources expected in H$_2$ dominated atmospheres, namely H$_2$O, CO, CH$_4$, CO$_2$, C$_2$H$_2$, HCN, NH$_3$, and H$_2$S, and collision-induced absorption (CIA) due to H$_2$-H$_2$ and H$_2$-He, obtained from the latest theoretical and experimental line databases currently available. 

Given a spectral dataset, the model is coupled with a parameter estimation method to obtain joint constraints on the chemical composition and temperature structure of the planet. Given a parametric temperature profile and molecular abundances, the model computes the emergent spectrum of the object at a high resolution which is then convolved with the instrument point spread function and binned to the data resolution for comparison with data, as discussed in section~\ref{sec:psf}. We explore the model parameter space and fit models to the data using a Markov Chain Monte Carlo (MCMC) algorithm \citep{madhusudhan2011a, line2015} and determine the posterior probability distributions of the model parameters. Given the very high SNR of the data we also inflate the uncertainties in the observed spectrum by $\sim$2\% of the maximum flux in the spectrum to account for any unaccounted systematics and/or model uncertainties \citep[see e.g.,][]{line2015}. The model space includes a wide range of plausible temperature profiles and compositions, spanning models with oxygen-rich as well as carbon-rich compositions. Thus, the model has 17 free parameters: six for the $P$-$T$ profile, eight for uniform mixing ratios of the eight molecules (H$_2$O, CO, CH$_4$, CO$_2$, C$_2$H$_2$, HCN, NH$_3$, and H$_2$S), and three parameters for the gravity, radius, and distance. This model set-up has been widely used to constrain chemical compositions and temperature profiles from infrared observations of dozens of giant exoplanetary atmospheres. 

\section{Results} 
\label{sec:results}

In this section, we report the constraints on the atmospheric compositions of the three objects in our sample. The atmospheric retrieval allows us to place joint constraints on the chemical abundances, the temperature profiles, and the macroscopic parameters such as gravity. The best fit model spectra to all the three objects are shown in Fig.~\ref{fig:spectrum}, and the estimated compositions and their posterior probability distributions are shown in Table~\ref{tab:results} and Fig.~\ref{fig:histograms}, respectively. 

The high-precision HST WFC3 spectra enable us to detect and derive constraints on the mixing ratios of key 
molecular species which, in turn, help constrain the corresponding elemental abundances. As discussed in section~\ref{sec:method}, we place constraints on the abundances of H$_2$O, CH$_4$ and  NH$_3$ which are known to have strong spectral signatures in the WFC3 bandpass (1.1-1.7 $\mu$m). For H$_2$-rich atmospheres, e.g., of giant planets and brown dwarfs, it is expected that O and C are largely contained in H$_2$O, CH$_4$, and CO, over a wide range of temperatures \citet{madhusudhan2012}. Furthermore, for temperatures below $\sim$1300 K in the observable atmospheres at pressures below $\sim$1 bar, H$_2$O and CH$_4$ contain all the O and C, respectively. Therefore, for the current objects in our sample which fall in this temperature range the elemental abundances of O and C can be derived from the mixing ratios of H$_2$O and CH$_4$, respectively. 

We report clear detections of H$_2$O, CH$_4$, and NH$_3$ in all the three objects. Figure~\ref{fig:histograms} shows the posterior probability distributions of the various molecular mixing ratios in the atmosphere. We do not detect any other molecule, e.g., CO or CO$_2$. As shown in Table~\ref{tab:results}, we measure the H$_2$O abundances (in $\log$(H$_2$O/H$_2$) of the three objects to be: -3.86$^{+0.10}_{-0.03}$, -4.49$^{+0.04}_{-0.12}$, and -4.30$^{+0.12}_{-0.04}$. The CH$_4$ abundances of the corresponding objects are -3.82$^{+0.14}_{-0.02}$, -4.71$^{+0.05}_{-0.11}$ and -4.66$^{+0.17}_{-0.03}$, respectively. Given the high S/N of the HST observations our constraints on the molecular abundances are able to achieve precisions below $\sim$0.15 dex, i.e. within a factor of 2, similar to that achieved by previous studies \citep[][]{line2014,line2015}. Even so, the precisions are limited not by the quality of the data but by the jitter added to the original observational uncertainties to account for unaccounted systematics and/or missing physics in the models. 

\begin{figure}[t]
\centering
\includegraphics[width = 0.5\textwidth]{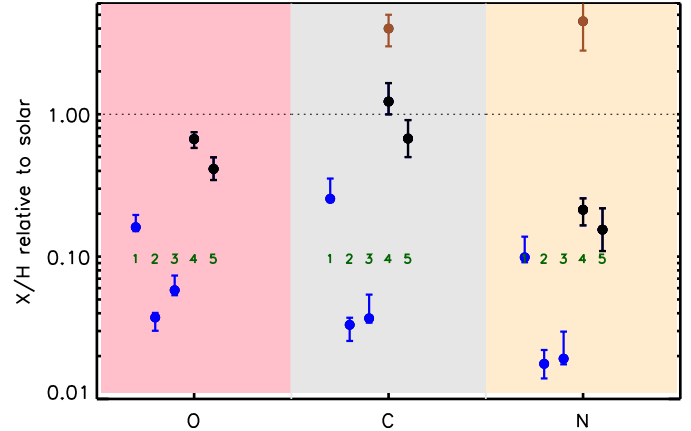} 
\caption{Atmospheric elemental abundances of the objects in our sample. The elemental abundances are derived from the estimated molecular abundances as discussed in section~\ref{sec:results}.} 
\label{fig:zplot} 
\end{figure}

We find sub-solar C/H and O/H abundances and diverse C/O ratios in our sample. The abundance estimates are shown in  Fig.~\ref{fig:zplot} and Table~\ref{tab:results}. As discussed above, the C/H and O/H ratios are derived from the CH$_4$ and H$_2$O mixing ratios, respectively. This is justified as given the low temperatures of these objects ($\lesssim$1000 K) CH$_4$ and H$_2$O are expected to be the dominant carriers of C and O, respectively, which is further supported by the fact that we do not detect any CO in these objects. We find the C/H and O/H ratios to be sub-solar in all the objects irrespective of their C/O ratios, which range between $\sim$0.35 and 1.0. Our derived abundances for the three systems follow a similar trend to that observed by \citet{line2015} in two other T Dwarfs. \citet{line2015} reported sub-solar O/H ratios for both their objects, while the C/H ratio was sub-solar for one and nearly solar (1-2$\times$solar) for the second object. 

We find a wide range of C/O ratios for our three objects, between oxygen-rich (C/O = 0.4) and carbon-rich (C/O = 1.0), while \citet{line2015} reported significantly carbon-rich ratios (C/O = 1-1.2) for both their objects. \citet{line2015} explained their low O/H ratios and high C/O ratios as possibly due to silicates, which are not observable,  taking up some of the oxygen thereby causing an apparent oxygen-poor (or carbon-rich) composition in an atmosphere that is actually oxygen-rich. We have, therefore, corrected our observed O/H abundances for this effect by allowing the possibility of up to 20\% of the oxygen locked in silicates; this number fraction of silicates is expected for a solar composition gas \citep{johnson2012,moses2013}. Even with that correction we still find that the O/H ratios are sub-solar and the C/O ratios span a wide range ($\sim$0.35-1.0), suggesting that the systems of Line et al may actually be carbon-rich. Overall,  the current sample of five objects, including ours and those from Line et al, suggest that all these objects are metal poor and span a diverse range of C/O ratios, both oxygen-rich and carbon-rich.  

We also find evidence for sub-solar N/H in all the T Dwarfs in our study. We derive the N/H abundance from the NH$_3$ mixing ratio. However, the N/H thus derived can also be considered as a lower limit since some of N can be bound in N$_2$ which does not have strong spectral features. Fig.~\ref{fig:zplot} shows the derived N/H abundances for the objects in our study as well as those from \citet{line2015} based on their NH$_3$ abundances. While one of our objects has a similar N/H ratio as the two objects of Line et al., the remaining two objects have significantly lower values. However, in all the five cases the derived N/H abundances are manifestly sub-solar, between 5-50$\times$ sub-solar. Given that the temperatures of all the objects are in a similar T Dwarf range it is not clear if the wide range of depletion can be explained solely by N being bound in unseen N$_2$. Moreover, the relative N/H abundances between the five objects follow the same trend as the O/H and C/H abundances in all the systems. This is less likely to be the case if N$_2$ contained most of the N with only trace amounts in NH$_3$. Therefore, this trend is suggestive of the NH$_3$ containing a significant fraction of N and the actual N/H abundances also being sub-solar like the O/H and C/H abundances. 

\begin{figure*}[t]
\centering
\includegraphics[width = 0.8\textwidth]{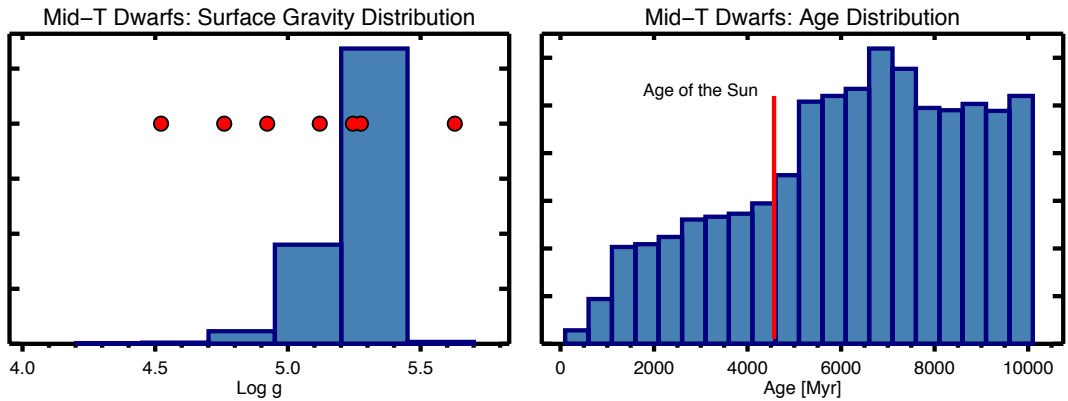} 
\caption{Distribution of ages and gravities of Mid-TDwarfs predicted by our fiducial evolutionary model assuming a flat star formation rate.} 
\label{fig:popsys1} 
\end{figure*}

\begin{figure*}
\centering
\includegraphics[width = 0.8\textwidth]{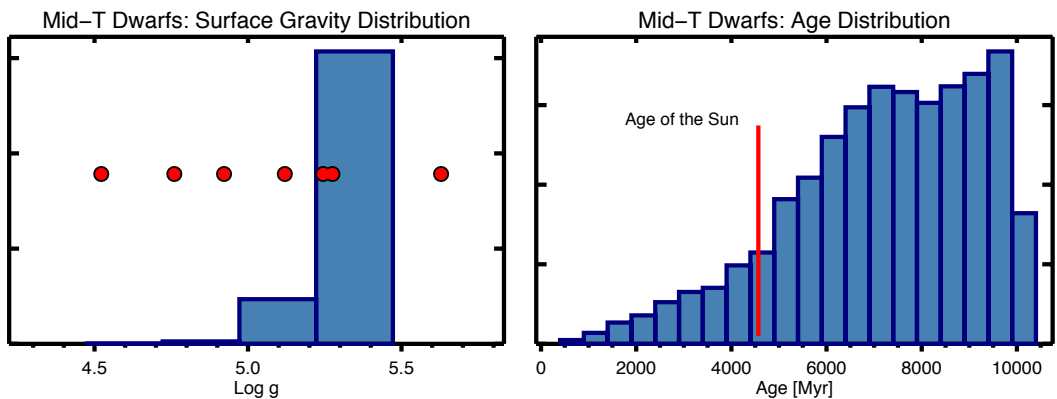} 
\caption{Distribution of ages and gravities of Mid-TDwarfs predicted by our fiducial evolutionary model assuming a  star formation rate that is linearly declining in time.} 
\label{fig:popsys2} 
\end{figure*}

\section{ Summary and Discussion}
\label{sec:discussion}

The atmospheric chemical abundances in the five T Dwarfs studied to date reveal a population very different in elemental composition from main-sequence stars in the solar neighborhood. For example, in four of the five objects we find oxygen metallicities below 0.5$\times$solar, i.e. [O/H] $<$ -0.3. In contrast, over 90\% of F, G, K main-sequence dwarf stars in the solar neighborhood have [O/H] $>$ -0.3 \citep{brewer2016}. Similarly, four of the five objects in our sample have C/O ratios greater than 0.7, whereas over 99\% of F, G, K stars have C/O $<$ 0.7. In particular, two of the objects in our current study have metallicities below 0.1$\times$ solar, i.e. [O/H] $<$ -1, which is rarely found in dwarf stars. Therefore, the predominantly sub-solar C and O abundances and the super-solar C/O ratios are incongruent with the abundance patterns observed in main-sequence stars in the solar neighborhood. Instead, such an abundance pattern leads to two interesting possibilities. Firstly, the abundance pattern is more reminiscent of the compositions of early generation stars which display significantly sub-solar metallicities and high C/Fe and/or C/O ratios \citep[see e.g.,][]{frebel2015,howes2015}. It might then be the case that the T Dwarfs in the sample have very old ages, are probes of the early galactic composition, and open a new avenue to galactic archeology. On the other hand, a second possibility is that the objects in our sample did not collapse from a proto-stellar cloud but instead formed as giant planets. We discuss both possibilities below. 

\subsection{Early Formation Epochs} 

The result that mid-T dwarfs in our sample often display sub-solar elemental abundances is unpredicted and may be surprising. We propose that the deduced sub-solar abundances stem from the old age of our targets, combined with the galaxy's chemical evolution (from sub-solar to super-solar).  This hypothesis may only be valid if the ages of our targets tend to significantly exceed that of the Sun. As brown dwarfs Ð unlike stars Ð lack stable hydrogen fusion, their luminosity evolution is strongly mass-dependent, making the determination of individual ages and masses difficult. 

Therefore, to explore the hypothesis we show the results of a toy model that mimics the formation, evolution, and selection of our targets.  In our model we combine a substellar initial mass function \citep{dario2012} and two  different brown dwarf formation rates (constant in time and linearly declining in time). Using brown dwarf evolutionary models based on radiative transfer and chemical equilibrium calculations \citep{baraffe2003}, we calculated the present-day temperature distribution of a simulated brown dwarf population. In the final step, we applied a temperature-based selection, emulating our choice of restricting our targets to T4ÐT8 spectral types. 

Figure~\ref{fig:popsys1} shows the predicted surface gravity and age distribution for a spectral type-selected brown dwarf sample assuming a constant brown dwarf formation rate.  Figure~\ref{fig:popsys2} shows the same predictions for a linearly declining brown dwarf formation rate. Both samples are dominated by brown dwarfs that are older than the Sun; in the latter sample Ð with star formation rate linearly declining Ð the ratio of older to younger brown dwarfs is even higher than in the former.

Both of these predictions are qualitatively consistent with the characteristically sub-solar abundances found in our analysis. We emphasize that the toy model presented here is not aiming to even approach the full complexity of the galactic chemical evolution, but it is merely a demonstration that the interplay of the substellar initial mass function and the gradual cooling of brown dwarfs tends to lead to old brown dwarfs dominating a sample of mid-T-type dwarfs. The comparison of the Figs. ~\ref{fig:popsys1} and ~\ref{fig:popsys2} demonstrates that our toy model is not sensitive to the input parameters; although varying the star formation rate does change the age distribution of the temperature-selected sample, it will not change the prediction that old brown dwarfs outweigh the young ones.

\subsection{Formation in Circumstellar Disks} 
The atmospheric chemical abundances of the five T Dwarfs provide initial insights into their possible formation conditions. The sub-solar metallicities (C, O, and N) and diverse C/O ratios observed in the objects ostensibly suggest a formation mechanism similar to giant planets formed in a disk. Firstly, the low metallicities are consistent with the natural trend in giant planets where the metallicities decrease with planetary mass. In the solar system, the metallicities, represented by  the C/H ratio, of the ice giants are $\sim$30-50$\times$solar, that of Saturn is $\sim$10$\times$ solar and that of Jupiter is 3-5$\times$ solar \citep{atreya2016}. An extrapolation of that trend would naturally suggest that brown dwarfs would have sub-stellar metallicities. Such a trend appears because the metallicity of a giant planet is governed by the amount of solids accreted during its formation relative to the amount of gas. Therefore for a gas giant, where most of the mass is in the gas, the lower the total mass the higher its atmospheric metallicity. The maximum amount of solids accreted by the object is limited by the available mass of solids in the disk. Thus, given their jovian or super-Jovian masses, brown dwarfs forming in a disk could be expected to posses very low metallicities. On the other hand, the C/O ratio of the object is governed by the formation location of the object relative to the snow lines of the major condensates \citep{oberg2011,madhusudhan2014c}; the C/O ratio of the gas increases with distance. Assuming predominant gas accretion, objects forming within the H$_2$O snow line can have solar C/O ratios, objects forming beyond the CO snow line can have C/O ratios of unity. Therefore, it may be the case that the objects we sample formed in a circumstellar disk akin to planets and, if found as isolated today, may have been ejected out of their systems by dynamical encounters. This possibility would be in line with the indication of a possible large population of planetary-mass brown dwarfs, or free floating giant planets, as inferred from the overabundance of short-timescale microlensing events \citep{sumi2011,clanton2016}.  

\acknowledgements{The observations used in this work are associated with HST program \#12550 and made with the NASA/ESA Hubble Space Telescope, obtained at the Space Telescope Science Institute, which is operated by the Association of Universities for Research in Astronomy, Inc., under NASA contract NAS 5-26555. We thank Esther Buenzli for sharing the data from her published work and for helpful discussions.} 
\vspace{5pt}

\bibliographystyle{apj}
\bibliography{references_bd}

\begin{thebibliography}{}
\expandafter\ifx\csname natexlab\endcsname\relax\def\natexlab#1{#1}\fi

\bibitem[{{Apai} {et~al.}(2013){Apai}, {Radigan}, {Buenzli}, {Burrows}, {Reid},
  \& {Jayawardhana}}]{apai2013}
{Apai}, D., {Radigan}, J., {Buenzli}, E., {et~al.} 2013, \apj, 768, 121

\bibitem[{{Atreya} {et~al.}(2016){Atreya}, {Crida}, {Guillot}, {Lunine},
  {Madhusudhan}, \& {Mousis}}]{atreya2016}
{Atreya}, S.~K., {Crida}, A., {Guillot}, T., {et~al.} 2016, ArXiv e-prints,
  arXiv:1606.04510

\bibitem[{{Baraffe} {et~al.}(2003){Baraffe}, {Chabrier}, {Barman}, {Allard}, \&
  {Hauschildt}}]{baraffe2003}
{Baraffe}, I., {Chabrier}, G., {Barman}, T.~S., {Allard}, F., \& {Hauschildt},
  P.~H. 2003, \aap, 402, 701

\bibitem[{{Barman} {et~al.}(2015){Barman}, {Konopacky}, {Macintosh}, \&
  {Marois}}]{barman2015}
{Barman}, T.~S., {Konopacky}, Q.~M., {Macintosh}, B., \& {Marois}, C. 2015,
  \apj, 804, 61

\bibitem[{{Brewer} {et~al.}(2016){Brewer}, {Fischer}, {Valenti}, \&
  {Piskunov}}]{brewer2016}
{Brewer}, J.~M., {Fischer}, D.~A., {Valenti}, J.~A., \& {Piskunov}, N. 2016,
  \apjs, 225, 32

\bibitem[{{Buenzli} {et~al.}(2014){Buenzli}, {Apai}, {Radigan}, {Reid}, \&
  {Flateau}}]{Buenzli2014}
{Buenzli}, E., {Apai}, D., {Radigan}, J., {Reid}, I.~N., \& {Flateau}, D. 2014,
  \apj, 782, 77

\bibitem[{{Buenzli} {et~al.}(2015){Buenzli}, {Saumon}, {Marley}, {Apai},
  {Radigan}, {Bedin}, {Reid}, \& {Morley}}]{Buenzli2015}
{Buenzli}, E., {Saumon}, D., {Marley}, M.~S., {et~al.} 2015, \apj, 798, 127

\bibitem[{{Buenzli} {et~al.}(2012){Buenzli}, {Apai}, {Morley}, {Flateau},
  {Showman}, {Burrows}, {Marley}, {Lewis}, \& {Reid}}]{Buenzli2012}
{Buenzli}, E., {Apai}, D., {Morley}, C.~V., {et~al.} 2012, \apjl, 760, L31

\bibitem[{{Buenzli} {et~al.}(2013){Buenzli}, {Apai}, {Radigan}, {Morley},
  {Burrows}, {Flateau}, {Showman}, {Marley}, {Reid}, {Lewis}, \&
  {Jayawardhana}}]{buenzli2013}
{Buenzli}, E., {Apai}, D., {Radigan}, J., {et~al.} 2013, in Protostars and
  Planets VI Posters

\bibitem[{{Burgasser} {et~al.}(2010){Burgasser}, {Cruz}, {Cushing}, {Gelino},
  {Looper}, {Faherty}, {Kirkpatrick}, \& {Reid}}]{burgasser2010}
{Burgasser}, A.~J., {Cruz}, K.~L., {Cushing}, M., {et~al.} 2010, \apj, 710,
  1142

\bibitem[{{Burgasser} {et~al.}(2006){Burgasser}, {Geballe}, {Leggett},
  {Kirkpatrick}, \& {Golimowski}}]{burgasser2006}
{Burgasser}, A.~J., {Geballe}, T.~R., {Leggett}, S.~K., {Kirkpatrick}, J.~D.,
  \& {Golimowski}, D.~A. 2006, \apj, 637, 1067

\bibitem[{{Burgasser} {et~al.}(2003){Burgasser}, {Kirkpatrick}, {Liebert}, \&
  {Burrows}}]{burgasser2003}
{Burgasser}, A.~J., {Kirkpatrick}, J.~D., {Liebert}, J., \& {Burrows}, A. 2003,
  \apj, 594, 510

\bibitem[{{Clanton} \& {Gaudi}(2016)}]{clanton2016}
{Clanton}, C., \& {Gaudi}, B.~S. 2016, ArXiv e-prints, arXiv:1609.04010

\bibitem[{{Da Rio} {et~al.}(2012){Da Rio}, {Robberto}, {Hillenbrand},
  {Henning}, \& {Stassun}}]{dario2012}
{Da Rio}, N., {Robberto}, M., {Hillenbrand}, L.~A., {Henning}, T., \&
  {Stassun}, K.~G. 2012, \apj, 748, 14

\bibitem[{{Frebel} \& {Norris}(2015)}]{frebel2015}
{Frebel}, A., \& {Norris}, J.~E. 2015, \araa, 53, 631

\bibitem[{{Helling} \& {Casewell}(2014)}]{helling2014rev}
{Helling}, C., \& {Casewell}, S. 2014, \aapr, 22, 80

\bibitem[{{Howes} {et~al.}(2015){Howes}, {Casey}, {Asplund}, {Keller}, {Yong},
  {Nataf}, {Poleski}, {Lind}, {Kobayashi}, {Owen}, {Ness}, {Bessell}, {da
  Costa}, {Schmidt}, {Tisserand}, {Udalski}, {Szyma{\'n}ski}, {Soszy{\'n}ski},
  {Pietrzy{\'n}ski}, {Ulaczyk}, {Wyrzykowski}, {Pietrukowicz}, {Skowron},
  {Koz{\l}owski}, \& {Mr{\'o}z}}]{howes2015}
{Howes}, L.~M., {Casey}, A.~R., {Asplund}, M., {et~al.} 2015, \nat, 527, 484

\bibitem[{{Johnson} {et~al.}(2012){Johnson}, {Mousis}, {Lunine}, \&
  {Madhusudhan}}]{johnson2012}
{Johnson}, T.~V., {Mousis}, O., {Lunine}, J.~I., \& {Madhusudhan}, N. 2012,
  \apj, 757, 192

\bibitem[{{Konopacky} {et~al.}(2013){Konopacky}, {Barman}, {Macintosh}, \&
  {Marois}}]{konopacky2013}
{Konopacky}, Q.~M., {Barman}, T.~S., {Macintosh}, B.~A., \& {Marois}, C. 2013,
  Science, 339, 1398

\bibitem[{{Krist} {et~al.}(2011){Krist}, {Hook}, \& {Stoehr}}]{TinyTim2011}
{Krist}, J.~E., {Hook}, R.~N., \& {Stoehr}, F. 2011, in \procspie, Vol. 8127,
  Optical Modeling and Performance Predictions V, 81270J

\bibitem[{{Line} {et~al.}(2014){Line}, {Fortney}, {Marley}, \&
  {Sorahana}}]{line2014}
{Line}, M.~R., {Fortney}, J.~J., {Marley}, M.~S., \& {Sorahana}, S. 2014, \apj,
  793, 33

\bibitem[{{Line} {et~al.}(2015){Line}, {Teske}, {Burningham}, {Fortney}, \&
  {Marley}}]{line2015}
{Line}, M.~R., {Teske}, J., {Burningham}, B., {Fortney}, J., \& {Marley}, M.
  2015, ArXiv e-prints, arXiv:1504.06670

\bibitem[{{MacKenty} {et~al.}(2010){MacKenty}, {Kimble}, {O'Connell}, \&
  {Townsend}}]{MacKenty2010}
{MacKenty}, J.~W., {Kimble}, R.~A., {O'Connell}, R.~W., \& {Townsend}, J.~A.
  2010, in \procspie, Vol. 7731, Space Telescopes and Instrumentation 2010:
  Optical, Infrared, and Millimeter Wave, 77310Z

\bibitem[{{Madhusudhan}(2012)}]{madhusudhan2012}
{Madhusudhan}, N. 2012, \apj, 758, 36

\bibitem[{{Madhusudhan} {et~al.}(2016){Madhusudhan}, {Ag{\'u}ndez}, {Moses}, \&
  {Hu}}]{madhusudhan2016}
{Madhusudhan}, N., {Ag{\'u}ndez}, M., {Moses}, J.~I., \& {Hu}, Y. 2016, \ssr,
  arXiv:1604.06092

\bibitem[{{Madhusudhan} {et~al.}(2014{\natexlab{a}}){Madhusudhan}, {Amin}, \&
  {Kennedy}}]{madhusudhan2014c}
{Madhusudhan}, N., {Amin}, M.~A., \& {Kennedy}, G.~M. 2014{\natexlab{a}},
  \apjl, 794, L12

\bibitem[{{Madhusudhan} {et~al.}(2011){Madhusudhan}, {Burrows}, \&
  {Currie}}]{madhusudhan2011a}
{Madhusudhan}, N., {Burrows}, A., \& {Currie}, T. 2011, \apj, 737, 34

\bibitem[{{Madhusudhan} {et~al.}(2014{\natexlab{b}}){Madhusudhan}, {Knutson},
  {Fortney}, \& {Barman}}]{madhusudhan2014a}
{Madhusudhan}, N., {Knutson}, H., {Fortney}, J.~J., \& {Barman}, T.
  2014{\natexlab{b}}, Protostars and Planets VI, 739

\bibitem[{{Madhusudhan} \& {Seager}(2009)}]{madhusudhan2009}
{Madhusudhan}, N., \& {Seager}, S. 2009, \apj, 707, 24

\bibitem[{{Moses} {et~al.}(2013){Moses}, {Madhusudhan}, {Visscher}, \&
  {Freedman}}]{moses2013}
{Moses}, J.~I., {Madhusudhan}, N., {Visscher}, C., \& {Freedman}, R.~S. 2013,
  \apj, 763, 25

\bibitem[{{{\"O}berg} {et~al.}(2011){{\"O}berg}, {Murray-Clay}, \&
  {Bergin}}]{oberg2011}
{{\"O}berg}, K.~I., {Murray-Clay}, R., \& {Bergin}, E.~A. 2011, \apjl, 743, L16

\bibitem[{{Sumi} {et~al.}(2011){Sumi}, {Kamiya}, {Bennett}, {Bond}, {Abe},
  {Botzler}, {Fukui}, {Furusawa}, {Hearnshaw}, {Itow}, {Kilmartin}, {Korpela},
  {Lin}, {Ling}, {Masuda}, {Matsubara}, {Miyake}, {Motomura}, {Muraki},
  {Nagaya}, {Nakamura}, {Ohnishi}, {Okumura}, {Perrott}, {Rattenbury}, {Saito},
  {Sako}, {Sullivan}, {Sweatman}, {Tristram}, {Udalski}, {Szyma{\'n}ski},
  {Kubiak}, {Pietrzy{\'n}ski}, {Poleski}, {Soszy{\'n}ski}, {Wyrzykowski},
  {Ulaczyk}, \& {Microlensing Observations in Astrophysics (MOA)
  Collaboration}}]{sumi2011}
{Sumi}, T., {Kamiya}, K., {Bennett}, D.~P., {et~al.} 2011, \nat, 473, 349

\bibitem[{{Yang} {et~al.}(2015){Yang}, {Apai}, {Marley}, {Saumon}, {Morley},
  {Buenzli}, {Artigau}, {Radigan}, {Metchev}, {Burgasser}, {Mohanty},
  {Lowrance}, {Showman}, {Karalidi}, {Flateau}, \& {Heinze}}]{Yang2015}
{Yang}, H., {Apai}, D., {Marley}, M.~S., {et~al.} 2015, \apjl, 798, L13

\bibitem[{{Yang} {et~al.}(2016){Yang}, {Apai}, {Marley}, {Karalidi}, {Flateau},
  {Showman}, {Metchev}, {Buenzli}, {Radigan}, {Artigau}, {Lowrance}, \&
  {Burgasser}}]{Yang2016}
---. 2016, \apj, 826, 8

\end{thebibliography}

\end{document}